\documentclass[aps,pra,twocolumn]{revtex4}
\usepackage{graphicx}
\usepackage{amsmath}
\usepackage{bm}
\def\paraH2{{\it p}-H$_2$}
\def\he4{He$^4$}
\def\Am2{\AA$^{-2}$}
\def\gapx{\lower 2pt \hbox{$\buildrel>\over{\scriptstyle{\sim}}$}}
\def\lapx{\lower 2pt \hbox{$\buildrel<\over{\scriptstyle{\sim}}$}}

\begin{document}

\title{Hard core repulsion and  supersolid cluster crystals
} 
\author{Massimo Boninsegni}
\affiliation{Department of Physics, University of Alberta, Edmonton, Alberta, Canada T6G 2E1}
\date{\today}
\begin{abstract}
We study  the effect of a short-ranged hard-core repulsion on the stability and superfluid properties of the cluster crystal phase of two-dimensional (2D) soft core bosons. Results of Quantum Monte Carlo simulations on a cogent test case suggest that the main physical properties of the phase remain unaltered if the  range $d$ of the inner repulsive core is sufficiently short,  even if the strength of the repulsion is  several orders of magnitude greater than the outer soft core barrier. Only if $d$ is an appreciable fraction of the size of the clusters ($\gapx$ 5\%) does  a sufficiently strong  hard core repulsion cause the crystal to break down into a homogeneous superfluid;
a moderate  inner core repulsion enhances  the superfluid response of the crystalline phase.
\end{abstract}
\maketitle

\section{Introduction}

One of the most exciting developments in cold atom physics is undoubtedly the theoretical possibility of synthesizing artificial phases of matter by tuning the interaction among atoms or molecules. This may pave the way to the observation of novel many-body phenomena, not (yet or easily) observed in naturally occurring condensed matter. A chief example is the putative supersolid phase of matter \cite{rmp}, 
long sought in the solid state system most likely to display it, i.e.,  a crystal of  $^4$He, but whose detection has so far eluded experimenters \cite{hallock}.
\\ \indent
After a decade of intense theoretical investigation, following the original claim of observation of non-classical rotational inertia in solid $^4$He by Kim and Chan \cite{kc} (a claim since recanted \cite{chan2}), the consensus seems to be that the most important physical  hinderance to supersolidity in condensed matter systems, is the strong repulsion that any interatomic or intermolecular pair interaction features at short distances. Such a repulsion originates from the Pauli exclusion principle, affecting the overlapping electronic clouds of different atoms or molecules as they are brought close together;  its range is typically comparable to the  lattice constant of the crystal, and is at the root of the thermodynamic instability of a dilute gas of point  defects such as vacancies or interstitials \cite{pollet07}, which according to  early theories of supersolidity might undergo Bose-Einstein Condensation \cite{Andreev69,Chester}. Indeed, there is strong numerical evidence that no supersolid phase occurs in the presence of a pairwise interaction potential featuring a repulsive core, even one whose growth at short distance is slow, like the Yukawa pair potential \cite{boninsegni11}. 
\\ \indent
On the other hand, it was suggested a long time ago that the superfluid ground state of an interacting Bose system could also 
feature a density modulation \cite{gross} (i.e., satisfy the definition of supersolid phase)  if the pair potential features a relatively ``flat" region at short inter-particle separation.  In  more recent times, Quantum Monte Carlo simulations have yielded robust evidence of a  supersolid    in the high density, low temperature region of the phase diagram of 2D soft core bosons \cite{saccani,saccani2}, as well as 
in that of other systems featuring similar pair-wise potentials \cite{boninsegni11}. Such a phase  consists of a self-assembled crystal of superfluid droplets (clusters); tunnelling of particles among adjacent droplets can establish phase coeherence and give rise to a superfluid transition at low temperature. 
\\ \indent
A many-body system with such a peculiar inter-particle interaction would  
have been regarded of merely academic interest until not so long ago, but it now appears as if it might be
realized experimentally in an assembly of cold Rydberg atoms \cite{ryd}. In particular, a physical mechanism known as  Rydberg blockade can give rise to a modified pair potential, precisely featuring the kind of  plateau at short distance that can underlie the supersolid cluster crystal phase described above \cite{henkel10}. \\ \indent 
It seems fair to state that, at this time,  Rydberg atoms constitute the most promising way of observing the supersolid phase of matter. In particular,  its unambiguous detection   is rendered possible by the direct imaging of the momentum distribution,  which displays Bragg peaks in correspondence of reciprocal lattice vectors \cite{greiner}. Although most of the theoretical predictions have been made for 2D systems, the physical arguments apply in three dimensions as well; in any case, it is possible to approach the 2D limit by confining particles spatially  using an external harmonic potential in the direction perpendicular to their motion (the so-called ``pancake" geometry).
\\ \indent
In view of such a concrete  possibility, a number of theoretical issues must be addressed, in order to guide in the design and interpretation of experiments aimed at carrying out the observation described above. In particular, it is important to assess the robustness of the predicted supersolid phase against details of the interaction over which one may not exercise (complete) control. Previous work \cite{boninsegni11}  has yielded convincing evidence that the behavior of the pair-wise interaction at long distances plays virtually no role in the appearance of the cluster crystal phase. It has been conjectured \cite{reatto}, that a {\it necessary} (not sufficient) condition for its occurrence, is that the Fourier transform of the two-body potential go negative (i.e., that the potential become attractive) in some range of $k$. This heuristic argument is consistent with all  the numerical evidence accumulated so far.
\\ \indent
Nothing quantitative is known about the effect of a hard core repulsion between particles at distances much shorter than the characteristic size of the clusters. As mentioned above, on very general grounds one knows that any physical interaction among atoms 
or molecules necessarily must feature 
such a repulsive core at sufficiently short distances. Thus, the experimental relevance of theoretical  predictions made for a  soft core model ultimately hinges on their robustness {\em vis-a-vis} such a short-distance repulsion. One can reasonably opine that, if a 
substantial difference in scales exists between the characteristic radius $R$ of the soft core of the potential (or, phrased more generally, of the region in which the repulsive part of the potential softens), and the (presumably much smaller) radius $d$ of the hard core, i.e., if $d/R<<1$, then the basic physics of the soft core system should be unaffected. Since the softening of the potential at short distances is tunable, by means of the Rydberg blockade, one may conclude that such a condition ought to be generally attainable.
\\ \indent
However,  even in the limit of vanishing radius (i.e., a $\delta$-function term), an additional  repulsive term of sufficient strength can render the Fourier transfor of the potential positive-definite, in principle undermining the conjectured condition of stability of the droplet crystal stated above.
Thus, it seems worthwhile to carry out a first principle numerical study aimed at assessing quantitatively the effect of a hard core repulsion at short distance, in order to gain greater quantitative understanding of the boundaries within which experiments aimed at observing the supersolid droplet crystal phase of soft core systems should be performed. 
\\ \indent
To this aim, we have performed Quantum Monte Carlo simulations of a 2D system of soft core bosons, with an additional hard core repulsion at short distances, also modeled as a simple rectangular barrier.
Specifically, we have studied the low temperature superfluid and structural properties of the system at a fixed density, and for a value of the soft core repulsion for which a supersolid cluster crystal phase is known to exist, on varying the strength and the radius of the inner hard core. Some of the results of our studies confirm our initial expectations,  whereas others seem nontrivial.
\\ \indent
Our main findings are the following:
\\ \indent
{\em a}) As the height of the inner barrier is increased, one generally first observes the rise of the superfluid response of the cluster crystal, caused by enhanced tunnelling of particles across adjacent clusters. For a sufficiently strong inner repulsion, the cluster crystal ``melts" into a homogeneous superfluid.
\\ \indent
{\em b}) The rise of the superfluid fraction with the height of the inner barrier is roughly linear, with a slope that increases monotonically with $d/R$. We could not establish a precise dependence based on the simulation data. For the lowest value of $d/R$ considered here (0.01),  the superfluid fraction remains constant, within the uncertainties of the calculation,  as the height of the barrier is varied over four orders of magnitude. This suggests that for a sufficiently short range, the cluster crystal phase may be stable against a hard core repulsion of arbitrary strength, even though the numerical simulations performed in this work do not allow us to make with confidence a mathematical statement of such breadth.
\\ \indent
The remainder of this paper is organized as follows: in Sec. \ref{mc} we introduce the model and provide computational details; in   Sec. \ref{sere} we illustrate our results and provide a theoretical interpretation. Finally, we outline our conclusions in Sec. \ref{conc}.
\section{Model and Calculation}\label {mc}
\begin{figure}[h]
\centerline{\includegraphics[height=2.0in]{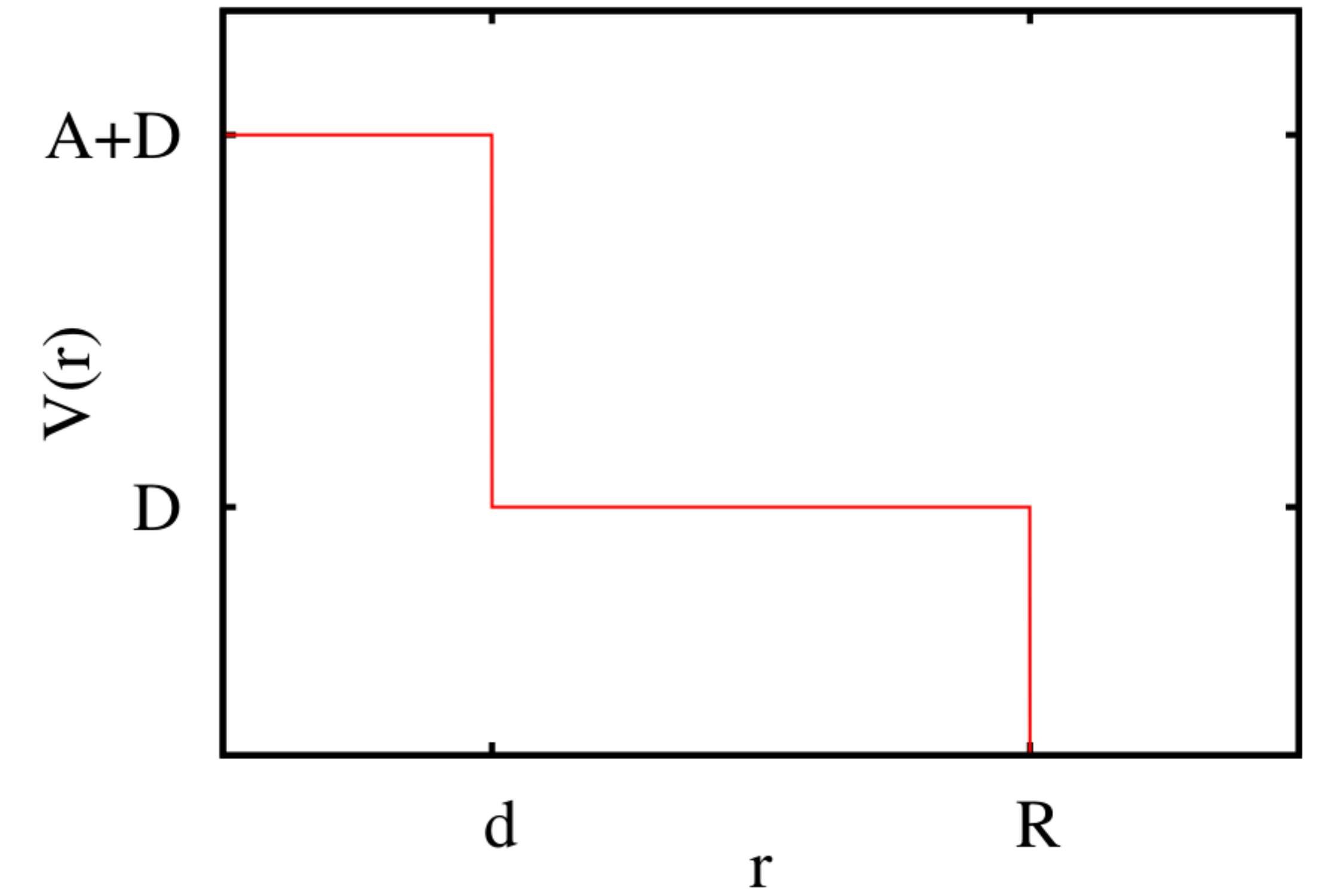}}
\caption{Schematic representation of the potential (\ref{pot}).}
\label{f0}
\end{figure}

Our system of interest consists of a collection of $N$ point-like particles of spin zero (hence obeying Bose statistics) and mass $m$, moving in two dimensions. The system is enclosed in a square simulation cell of side $L$, with periodic boundary conditions in all directions.  The density is $\theta=N/L^2$. Particles are assumed to interact via the following pair potential, only depending on the inter-particle distance:
\begin{equation}\label{pot}
V(r)= A\ \Theta(d-r)+D\ \Theta(R-r)
\end{equation}
where $\Theta(x)=1$ if $x>0$, zero otherwise, $d <R$ are the inner and outer radii,  and $A,\  D > 0$. $D$ is the height of the potential barrier at the outer radius $R$, whereas $A+D$ is that at the inner core radius $d$. If $A=0$, then (\ref{pot}) represents a soft core potential of radius $R$. Henceforth, we  take $\epsilon=\hbar^2/mR^2$ as our energy and temperature unit, and $R$ as the unit of length. The potential (\ref{pot}) is schematically shown in Fig. \ref{f0}. 
 \indent 

While obviously oversimplified, such a  minimal model contains all the necessary elements allowing us to address the physical issues of interest. The basic qualitative and quantitative results are not expected to depend significantly on the details of the potential at the two most relevant distances, namely $d$ and $R$ (this conclusion has already been established quantitatively for the case of soft core repulsions \cite{boninsegni11}).
\\ \indent
The protocol adopted here to investigate the low temperature properties of the system described above, consists of selecting first values of the outer (soft) core barrier height $D$ and of the particle density $\theta$ for which a supersolid cluster crystal phase exists  at low temperature  for $A=0$, namely with no inner hard core \cite{boninsegni11}. For every such choice of $D$ and $\theta$, we then obtained results for different parameters of the inner core, i.e., $d$ and $A$ in Eq. (\ref{pot}). 
The results presented here only pertain to one specific choice of $D$ ($D=5$) and $\theta$ ($\theta=4$), but the same qualititative trend is observed for other choices.
\\ \indent Our study consists of Quantum Monte Carlo simulations, based on  the Worm
Algorithm in the continuous-space path integral representation \cite{worm,worm2}. Because this well-established computational methodology is thoroughly described elsewhere, we do not review it here. The most important aspects to be emphasized here, are that it enables one to compute thermodynamic properties of Bose systems at finite temperature, directly from the microscopic Hamiltonian, in particular  energetic, structural and superfluid properties, in practice with no approximation. 
In particular, it has consistently proven a superior option in the investigation of the low temperature physics of Bose systems, featuring significant advantages over ground state methods such as Diffusion Monte Carlo (DMC), the most notable being the absence of bias arising from the use of a trial wave function or a finite population of random walkers \cite{bm,h2}, as well as the fact that it offers access to off-diagonal correlations such as the one-body density matrix, not accessible exactly to DMC \cite{mb}.
\\ \indent
Technical details of the simulation are standard, and we refer the interested reader to Ref. \onlinecite{worm2}. 
We used the primitive high-temperature approximation for the many-particle propagator; while this is not the most efficient form for the particular system of interest, nonetheless it proves adequate for the scope of this study. All of the results reported here are extrapolated to the limit of vanishing imaginary time 
step $\tau$. We considered systems comprising a number of particles $N$=144 and 576, obtaining consistent results for the two system sizes.
\\ \indent
The relevant physical quantities computed in this study are the pair correlation function, which allows one to monitor structural changes (away from the cluster crystal phase), as well as the superfluid density, both the global (computed using the well-known ``winding number" estimator \cite{pollock}), as well as the local one (i.e., inside the individual  clusters \cite{noi}).  Finally, the Worm Algorithm also allows for the unbiased evaluation of  the one-body density matrix, which is connected to the observable momentum distribution.

\section{Results}\label{sere}

\begin{figure}[h]
\centerline{\includegraphics[height=3.2in]{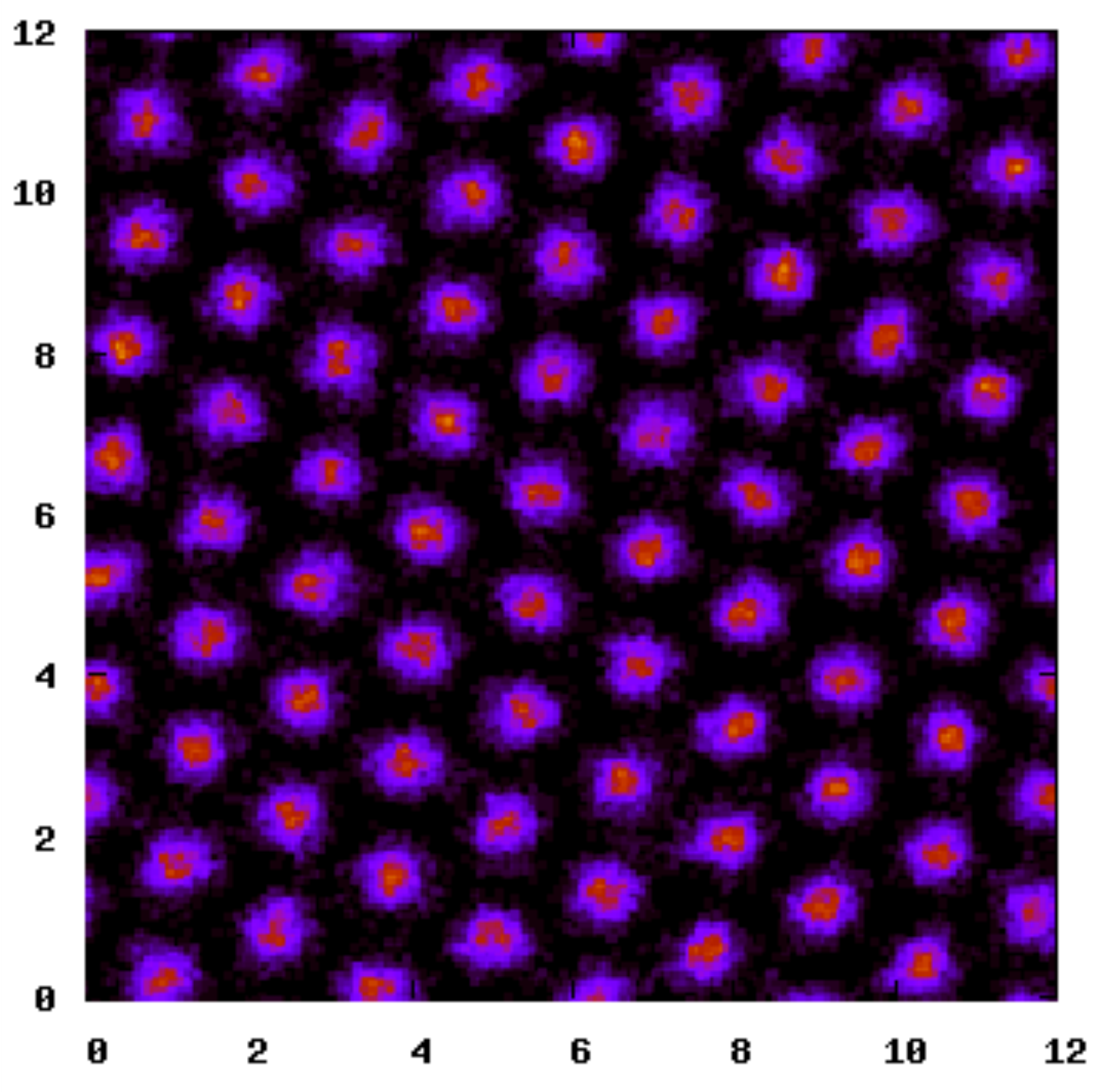}}
\caption{Density snapshot of a soft core system with $D$=5 at a density $\theta=4$ and at temperature $T$=0.5 (see text for the units). The simulated system comprises 576 particles. }
\label{f1}
\end{figure}

\begin{figure}[h]
\centerline{\includegraphics[height=2.3in]{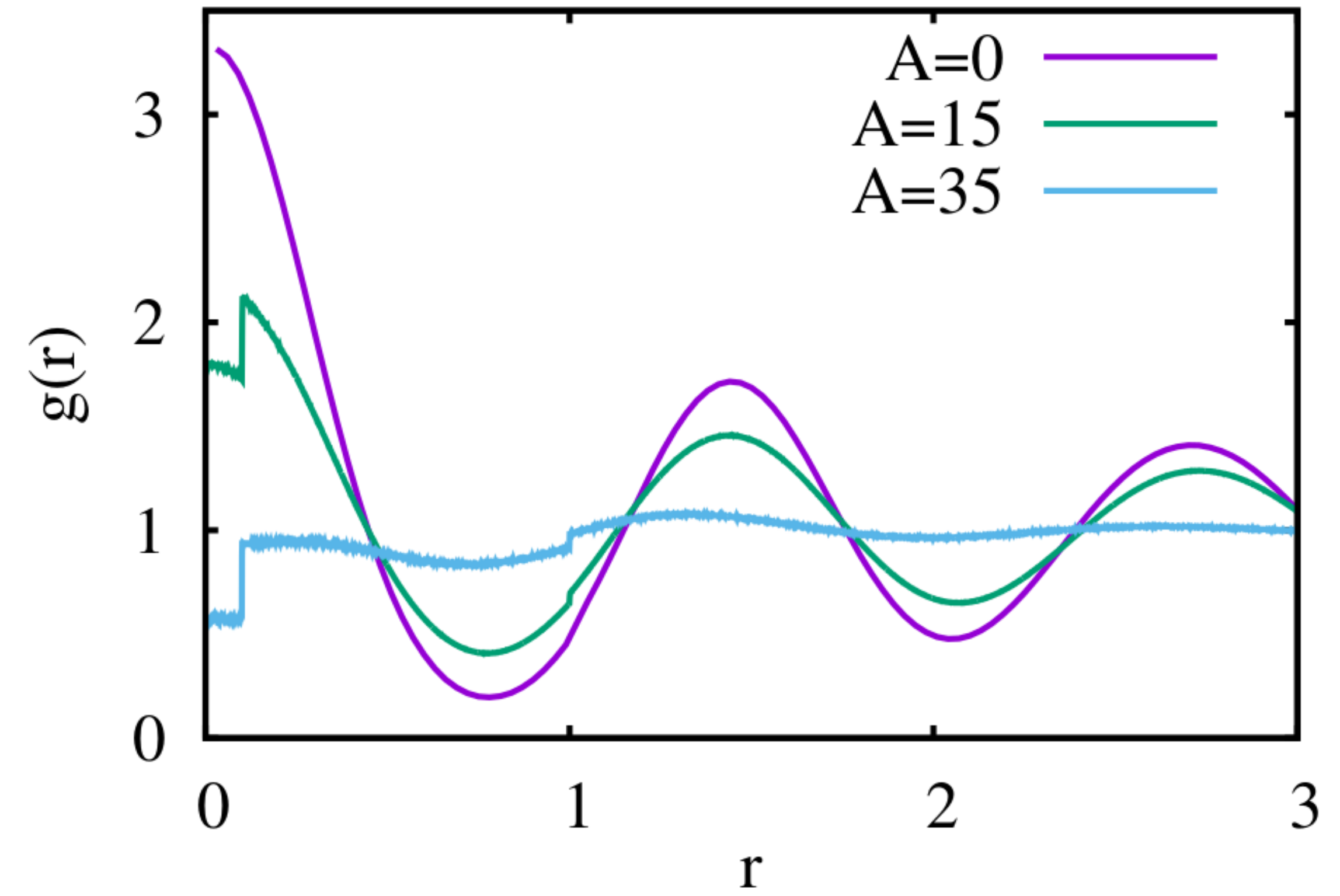}}
\caption{Pair correlation function for the system at temperature $T$=0.5, computed for three different values of the parameter $A$ in Eq. \ref{pot}. The value of the inner core radius $d$ is 0.1.}
\label{f3}
\end{figure}

\begin{figure}[h]
\centerline{\includegraphics[height=2.3in]{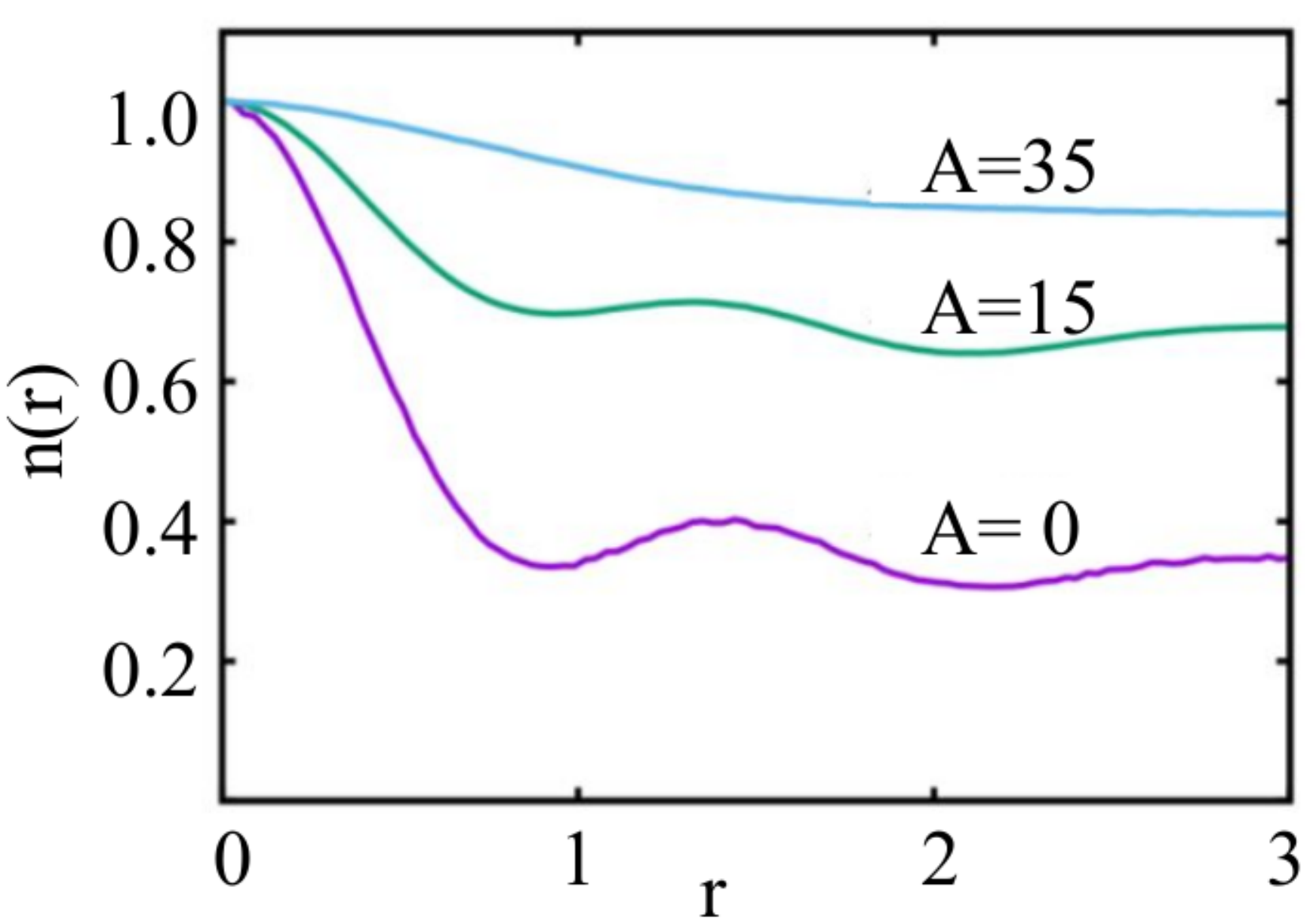}}
\caption{One-body density matrix computed at temperature $T$=0.5, computed for three different values of the parameter $A$ in Eq. \ref{pot}. The value of the inner core radius $d$ is 0.1.}
\label{f4}
\end{figure}

If $A=0$ in (\ref{pot}), the density $\theta=4$ (i.e., the mean interparticle distance is 0.5 $R$) and $D=5$, then the ground state of the system is a supersolid cluster crystal,  with clusters comprising approximately 7 particles arranged on a triangular lattice (shown in Fig. \ref{f1}). The superfluid fraction of the system is $\sim $ 8\% in the $T\to 0$ limit \cite{boninsegni11}. 
\begin{figure}[h]
\centerline{\includegraphics[height=2.5in]{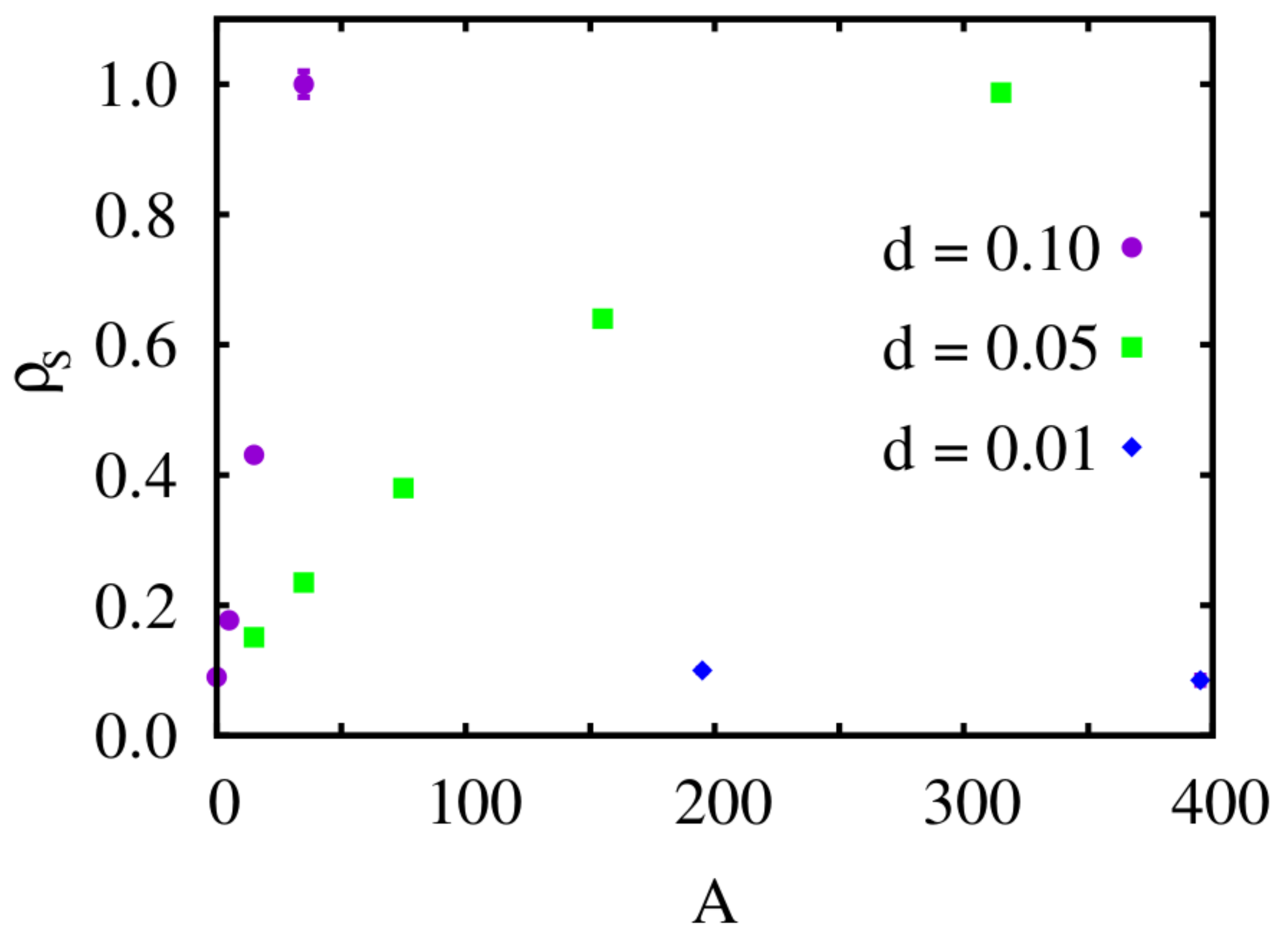}}
\caption{Superfluid fraction $\rho_S$ of the system at temperature $T$=0.5, computed as a function of the parameter $A$ of the inner hard core of the interparticle potential (Eq. \ref {pot}), for different values of the inner core radius $d$.
When not explicitly shown, statistical errors are smaller of the symbol size.}
\label{f2}
\end{figure}
Physically, the effect of a hard core repulsion at short distances can be expected to be that of breaking down the clusters, thereby restoring translational invariance and stabilizing a homogeneous superfluid.  
\\ \indent
Fig. \ref{f2} shows the computed superfluid fraction $\rho_S$ of the system as a function of the parameter $A$ of the inner core of the interparticle potential (Eq. \ref{pot}), for three different values of the inner core radius $d$ (expressed in units of $R$, the outer, ``soft" core radius). All of the results  are at temperature $T$=0.5 (in our energy units),   empirically verified to be low enough for the simulated system 
to be essentially in its ground state \cite{zz}. In particular, the two data points for which the superfluid fraction is close to unity correspond to a physical situation in which the system is   a uniform superfluid in the $T\to 0$ limit, whereas in all other cases a supersolid cluster crystal occurs \cite{note}. 
\\ \indent
For the  largest value of $d$ (0.1),  $\rho_S$ increases rapidly with $A$, approaching unity for $A\sim 35$. However, for intermediate values of $A$ (e.g., 15), there is evidence that the system is still in the supersolid phase, as shown by the {\it a}) well-defined oscillations of the pair correlation function $g(r)$, shown  in Fig. \ref{f3}, as well as by density snapshots such as the one shown in Fig. \ref{f1}, and {\it b}) the fact that the superfluid fraction saturates to a value less than unity in the $T\to 0$ limit, with no appreciable dependence on the system size.  
As $A$ is increased further, to 35, the $g(r)$ is essentially flat, and any remnant of solid order has disappeared; concurrently, the superfluid fraction is essentially 100\%.
\\ \indent
The superfluid fraction for $A=15$ is close to 40\%, i.e., significantly strengthened with respect to the soft core 
(i.e., $A=0$) case. The physical reason for the increased superfluid response is the enhanced tunnelling between adjacent clusters caused by the higher inner potential barrier. This is confirmed by the calculation of the one-body density matrix $n(r)$, shown in Fig. \ref{f4}, always at $T$=0.5 and for $d=0.1$, for the same three values of $A$ of Fig. \ref{f3}. The oscillations which are a signature of the supersolid are still clearly visible for $A$=15, whereas no evidence of them is left for $A$=35.
The general trend illustrated by these results can also be observed for $d$=0.05. The main difference is that the solid order is more resilient, i.e., a greater height of the inner barrier is needed in order to ``melt" the crystal. For example, in this case the system still displays solid order at this temperature for $A\sim 30$, with a value of $\rho_S$ close to as much as 70\%. All of this points to a gradual change of the nature of the ground state, namely the presence of multiple barriers of increasing (decreasing) height (radius) causes on the one hand the weakening of the crystalline order, on the other an enhancement of the superfluid response; however, both types of order coexist in a rather extended range of $A$.  Fig. \ref{f1} shows a  linear dependence of $\rho_S$ on $A$ in the supersolid phase, for $d=0.1$ and 0.05, with a slope that increases monotonically with $d/R$ (the results are not consistent with a dimensionally expected $(d/R)^2$ dependence). 
\\ \indent
We were unable to fit the results obtained at the three different values of $d$ into a single pattern. As shown in Fig. \ref {f2}, the estimates of the superfluid  fraction for the smaller value of $d$ considered, ($d=0.01$), are within the statistical errors of the calculation independent of the parameter $A$, at least over a range spanning several orders of magnitude. Indeed, we could not detect any significant change in the  value of $\rho_S$, with respect to that corresponding to $A=0$, for $A$ as large as $2\times 10^4$ \cite{pipo}. Within the obvious limitations that a numerical study such as this one undoubtedly features, when it comes to making rigorous mathematical statements, these results do not rule out that, in the $d\to 0$ limit, the supersolid cluster crystal phase may be robust against short-range, hard core repulsion of arbitrary strength. This would undermine the validity of the heuristic criterion \cite{reatto} according to which the presence of a negative Fourier component in the interaction potential would be a necesary condition for the occurrence of a cluster crystal. Obviously, further studies will be needed to come to a definite conclusion on this subject. 
\\ \indent
\section{Conclusions}\label{conc}
We have studied by means of Quantum Monte Carlo simulations the stability of the 2D supersolid cluster crystal phase  of soft core bosons, in the presence of an inner repulsive hard core, which necessarily must exist in any realistic interatomic or intermolaecular pair-wise interaction. In general, within the obvious oversimplification of the model,  the results point to the robustness of the phase, if there exists a substantial difference in scale between the spatial extent of the region in which the potential is slowly varying (``flat"), and the radius of the hard core repulsion. Indeed, our numerical results for a representative test case suggest that, if the ratio between inner and outer core radii is of order 1\% or less, then the supersolid cluster crystal phase may be stable for an arbitrarily high inner barrier.
\\ \indent
On the other hand, for $d/R\ \gapx\ 0.05$ a sufficiently high inner barrier causes the droplet crystal phase to melt into a homogeneous superfluid; there exists an intermediate range of inner core repulsive strength, however, the supersolid phase remains stable and the superfluid response is actually enhanced, as a result of increased tunnelling of particles among adjacent clusters. One could therefore think of exploiting this mechanism by engineering a modified soft core potential, featuring two or more steps; whether that is presently feasible using the Rydberg blockade or some other mechanism is unknown to us, but seems worthwhile a scenario to explore, as enhancing the signal will likely result in a more straightforward and unambiguous experimental detection.

\section*{Acknowledgements}

This work was supported  by the Natural Science
and Engineering Research Council of Canada. Computing support of Westgrid is gratefully acknowledged.


\begin{thebibliography}{0}
\bibitem{note}
It should be noted that because the system breaks translational invariance in the ground state, the superfluid fraction does {\it not} saturate to 100\% in the $T\to 0$ limit.
\end{thebibliography}
\end{document}